\documentclass[preprint,aps]{revtex4}
\usepackage{amsmath, amsthm, graphicx, amssymb,url}
\usepackage{verbatim}
\usepackage[usenames, dvipsnames]{color}

\usepackage[ pdftex, plainpages = false, pdfpagelabels, 
                 pdfpagelayout = useoutlines,
                 bookmarks,
                 bookmarksopen = true,
                 bookmarksnumbered = true,
                 breaklinks = true,
                 linktocpage=all,
                 pagebackref=false,
                 colorlinks = true,
                 linkcolor = BrickRed,
                 urlcolor  = blue,
                 citecolor = BrickRed,
                 anchorcolor = green,
                 hyperindex = true,
                 hyperfigures]{hyperref}
\usepackage{epstopdf}

\usepackage{dpfloat}

\newcounter{figref}	

\newcommand{\vect}[1]{\vec{#1}}
\newcommand{\caprule}{\hspace{-15mm}\rule[3.5mm]{55mm}{.5pt}\hspace{-40mm}\rule[3.45mm]{25mm}{.9pt}}

\begin{document}

\title{\href{http://necsi.edu/research/social/afghanag.html}{Modeling Policy and Agricultural Decisions in Afghanistan}\footnotetext{A revised version of this manuscript has been published after peer review in GeoJournal: \\\url{http://www.springerlink.com/openurl.asp?genre=article&id=doi:10.1007/s10708-012-9453-y}}}
\author{Michael J. Widener,$^{1,2}$ Yavni Bar-Yam,$^{1}$ Andreas Gros,$^{1}$  \\ Sara Metcalf,$^{2}$ and \href{http://necsi.edu/faculty/bar-yam.html}{Yaneer Bar-Yam}.$^{1}$}
\affiliation{{$^1$} \href{http://necsi.edu}{New England Complex Systems Institute}, 238 Main St., Suite 319, Cambridge, MA 02142, USA \\
{$^2$} \href{http://www.geog.buffalo.edu/}{Department of Geography, University at Buffalo}, 105 Wilkeson Quad, Buffalo, NY 14261, USA}

\begin{abstract}
Afghanistan is responsible for the majority of the world's supply of poppy crops, which are often used to produce illegal narcotics like heroin. This paper presents an agent-based model that simulates policy scenarios to characterize how the production of poppy can be dampened and replaced with licit crops over time. The model is initialized with spatial data, including transportation network and satellite-derived land use data. Parameters representing national subsidies, insurgent influence, and trafficking blockades are varied to represent different conditions that might encourage or discourage poppy agriculture. Our model shows that boundary-level interventions, such as targeted trafficking blockades at border locations, are critical in reducing the attractiveness of growing this illicit crop. The principle of least effort implies that interventions decrease to a minimal non-regressive point, leading to the prediction that increases in insurgency or other changes are likely to lead to worsening conditions, and improvements require substantial jumps in intervention resources.

\end{abstract}

\maketitle

\setlength{\parskip}{1ex}

\section{Introduction}

The conflict between the myriad political actors in Afghanistan over the past decade has resulted in a tremendous amount of social and economic instability. The inability to establish an adequate governance structure within the nation has adversely affected the integration of Afghanistan into the global political and economic system \citep{BarfieldNojumi2010,Shroder2007}. One outcome of this tumultuous socio-political environment has been the reemergence of poppy farming, particularly in the southern and eastern portions of the country \citep{Goodhand2000,Goodhand2005}. Previously, during the Soviet invasion in 1978, Afghani warlords utilized the production of poppy to fund their regional armies \citep{CramerGoodhand2002}. Similarly, since 2002, the cultivation and sale of poppy plants has played an important role in financing anti-NATO and -U.S. resistance forces. Poppy farming and its related activities hinder efforts to establish more legitimate economies throughout poppy growing regions.

Peters argues that the poppy trade has greatly affected the motives of al Qaeda in Afghanistan and the Taliban, noting that some incursions are designed to protect drug shipments rather than advance political and territorial goals \citep{Peters2009book}. Additionally, many have written that political corruption is a problem tightly related to the opium trade \citep{Gannon2004,Peters2009book,Goodhand2008,Blanchard2009}. Peters and Goodhand in particular argue that current zero-tolerance strategies, which are a result of poppy criminalization, are counterproductive. Understanding and dealing with the poppy trade in a way that does not victimize less prominent actors in the system is an important yet complicated component of bringing peace to Afghanistan.

A group of ``less prominent actors'' of particular interest in this system are Afghan farmers. Farmer decisions to cultivate this crop are primarily based on economics. In their 2010 Afghanistan Opium Survey, the United Nations Office on Drugs and Crime \citep{UNODC2010} reports that poverty and the cropÕs expected high sale price are the primary reasons farmers choose to grow poppy \citep{UNODC2010a}. However, this decision also takes place in a complex system of dangers and rewards, government policy, political motivations, insurgency, local infrastructure, and other factors, all of which indirectly influence the economic viability of poppy cultivation. Aside from the market price received for poppy, intermediaries may use non-economic forces as leverage for increasing or decreasing the benefits farmers receive from the poppy trade. For example, some Afghan farmers have reported that their ``decision to plant opium [has] been `influenced' by local commanders'' \citep{UNODC2004}. Another example of the complex situation on the ground relates to the fact that fertilizer is banned in Afghanistan due to its frequent use in improvised explosive devices (IEDs), perhaps making the hardy poppy plant an attractive crop choice \citep{Jaffe2011}.

Here we characterize the elements influencing poppy growth in Afghanistan, encapsulating both the economic and non-economic incentives into a single measure in a dynamic simulation that systematically summarizes the various economic and non-economic factors as dollar amounts in order to gain a better understanding of why farmers decide to grow poppy. We construct an agent-based model of Afghan farmers' crop decisions over time and across space, in order to explore how these decisions might change given different policy scenarios, expressed by key leverage parameters in the model. The model is calibrated with actual crop prices and satellite derived land-use imagery. We find that increases in the levels of insurgency significantly increase the level of poppy production in the regions in which it already exists, as well as in other regions that are proximate to unsecured border crossings. On the other hand, policies of increased national subsidy for licit crops and increased security on the Afghanistan--Pakistan border are effective in transitioning regions of poppy growth to licit crops.

An understanding of why farmers choose to grow certain crops is necessary to inform policy decisions aimed at establishing a more sustainable national economy. Identifying effective interventions is both a national and international concern, as 93\% of global opiate production currently occurs in Afghanistan \citep{Clemens2008,Goodhand2008}. Another critical benefit from an effective poppy reduction policy would be a parallel reduction in drug-related corruption within the Afghan government \citep{Glaze2007}. While exogenous forces, such as world demand for heroin, directly influence the dynamics of poppy production, a study of endogenous factors, such as local government and insurgency policies, can be a basis for local interventions. Interventions should recognize the need to provide alternative economic opportunities, as the current poppy growth plays an important economic role in the society as well as in individual lives---another reason policy directed at economic incentives could be both more robust and more generally beneficial to the society. Rubin \citep{Rubin2007} notes that past U.S.-led counter-narcotic measures have been used by the Taliban to gain support against international forces. By complementing or replacing these measures with economically supportive alternatives, Taliban forces will have less room to exert their own influence. The next section describes the structure of the simulation. We construct a model of empirically estimated economic forces and address questions whose answers are robust to the uncertainties of the estimates. The third section describes the outcomes from a number of scenarios. A summary of implications is given in the concluding section.

\section{Model Structure and Data}

The dynamic model of agricultural choice is composed of a number of independent farmer agents, who choose what type of crop to grow on their land. {It is assumed that the agents use farming methods appropriate to their region (e.g. use of irrigation or dryland farming).} Decisions {of which crop to grow} are based on a comparison of the net compensation they expect to receive from growing a licit crop or poppy, including social and logistical factors. {As is understood in the general theory of market function, the role of prices in our model serves as an aggregator of many complicated details of agricultural and social processes.} The value for licit crops is given by:
\begin{equation*}
v_{c}(t,\vect{r})=p_{c}+s(t,\vect{r})-i_{c}(1-\delta_{\,c(t-1,\vect{r})\,,\,c})
\end{equation*}
where the index $c$ takes the values grain-cereal-fodder or fruit-nuts-vegetables, for the two different types of licit crops grown by the farmer agents in this model. Here $p_{c}$ is the price a farmer receives for selling their harvest, $s$ represents the subsidies received from national and international organizations (including indirect benefits like equipment and improved distribution capacities), and $i_{c}$ is a one time cost of switching to a licit crop if they were growing poppy during the previous season. Here $c(t,\vect{r})$ is the crop raised at time $t$ at location $\vect{r}$. The Kronecker delta function, $\delta_{i,j}$ is equal to one when its two subscripts are equal ($i=j$) and zero otherwise ($i \neq j$), functioning as a switch to ``turn on'' the initiation cost term only if the farmer agent (at location $\vect{r}$) has not been growing the licit crop. The value for poppy is determined by:
\begin{equation*}
v_{poppy}(t,\vect{r})=p_{poppy}+f(\vect{r})-e(\vect{r})-i_{poppy}(1-\delta_{\,c(t-1,\vec{r})\,,\,poppy})
\end{equation*}
where $p_{poppy}$ is the price a farmer receives for selling their harvest. The force exerted on the farmer by insurgency, $f$, includes indirect benefits and incentives as well as the avoidance of strong-arm tactics. This is given by $f=f_0+f_1(\vect{r})$, where $f_0$ applies to all of Afghanistan and is a parameter adjusted in the model, and term $f_1(\vect{r})$ is particular to each province (and therefore spatially dependent). The cost of trafficking the drugs to the nearest exit point from the country, $e$ depends on the distance from the centroid of the province to the exit point, via the road network, and thus depends on the location $\vect{r}$. The ``one time'' initiation cost $i_{poppy}$ is the cost of switching from a licit crop to poppy if the farmer at the given location was growing a licit crop during the previous season.
At each time step (representative of the annual growing season) a farmer compares $v_{poppy}$ with $v_{licit}$ and chooses to grow the crop with the greatest value. The initiation costs $i_c$ are included to account for any additional labor, equipment or other overhead that might be incurred in switching a field from one crop to another. It embodies the economic value of the persistence of a farmer's growing practices.

\begin{figure*}[tb]
\refstepcounter{figref}\label{initial}
\centerline{
\includegraphics[width=150mm]{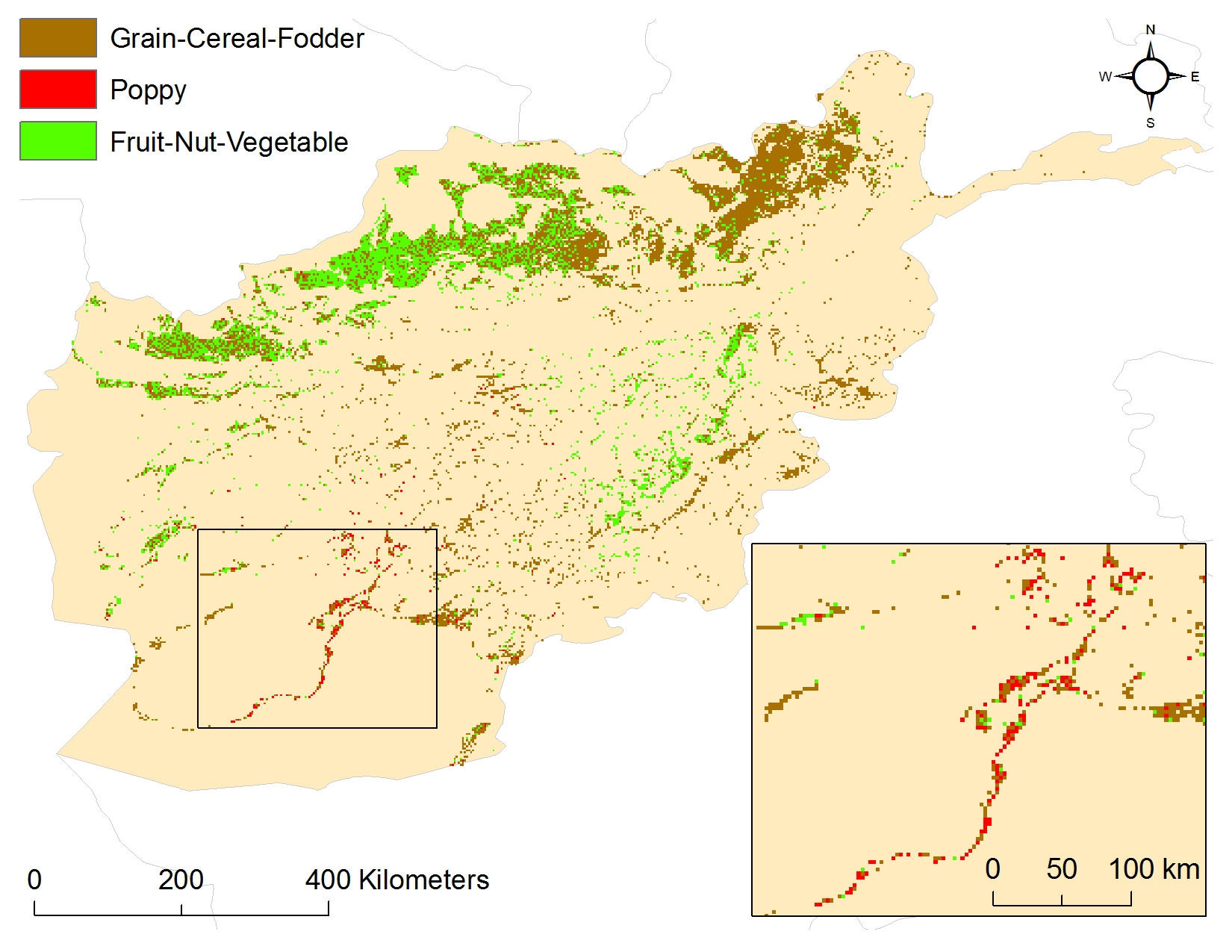}
}
\caption{Initial agricultural state: Crop cultivation initial conditions for three crop types within the area of Afghanistan: fruit-nut-vegetable (green) grain-cereal-fodder (brown), poppy (red) and uncultivated, within Afghanistan (off-white).}
\caprule
\end{figure*}

Land use data were obtained from Afghanistan Information Management Services and all arable land (e.g. land designated as suitable for irrigated, rain-fed, or vineyard agriculture) was identified to create a grid with 0.025 degree-squared cells, approximately equivalent to an area of 1.75 miles squared. Each cell was associated with the province in which it is located, and assigned one of two licit crop types (grain-cereal-fodder or fruit-nuts-vegetables) based on information from 2008 reports prepared by the National Agricultural Information System for the Afghan Ministry of Agriculture, Irrigation, and Livestock \citep{Bell2008} or, in cases where these reports are not available, from 2007 reports from the United Nations' World Food Programme's Food Security Atlas \citep{WFP}. A distinction between the aggregated categories of grain-cereal-fodder and fruit-nut-vegetable is made due to their different environmental requirements and sale prices. Information on the number of hectares of poppy grown in each province \citep{UNODC2010} was used to assign some cells as poppy growing regions. The result of this process is shown in Figure \ref{initial}.  In the model each farmer remembers its initial base designation (grain-cereal-fodder or fruit-nut-vegetable) and only adopts its base crop type or poppy throughout the simulation. 

The initial area (approximately 12,500 cells or 5,700,000 hectares) and distribution of crops corresponds well with published studies' documentation of where various types of crops are located in the country and their volume of production \citep{EuropeanCommission2006,UNODC2010}.  In particular, the initial state assigns approximately 275 cells to poppy cultivation, which is equivalent to around 125,000 hectare, consistent with recent UNODC reports \citep{UNODC2009,UNODC2010a}. There is slight variation in this value between different simulations of the model because crop assignment is probabilistic. 
For the purposes of our analysis it is assumed that prices for all crops remain constant, with grain-cereal-fodder crops priced at \$1,000 per Ha \citep{UNODC2010}, fruit-nut-vegetable crops at \$9,000 per Ha \citep{Department2010}, and poppy at \$3,750 per Ha \citep{UNODC2009}. Available subsidy per hectare ($s$), insurgent influence ($f_{gen}$), and cost of trafficking ($e$) are all parameters that are directly or indirectly varied to represent distinct conditions and policy interventions. 
The space and time dependence of $s(t,\vect{r})$ reflects the variation of conditions of individual farmers, even from plot to plot in the same neighborhood and over time. We incorporate a local random variation in the model by varying the effective subsidy a farmer receives with uniform probability in the range $\pm$ 10\% of the subsidy parameter value. The dynamics of the model arises from the possibility of time varying parameters and the effect of the random conditions.

The local insurgent influence parameter, $f_1(\vect{r})$, is assigned an initial value using a map from the International Council on Security and Development, depicting the level of insurgent activity in Afghan provinces \citep{ICOS2009}. The influence parameter uses a dollar amount to represent the added financial incentive (or avoidance of financial loss) faced by farmers confronted with insurgent forces. {In addition, direct threats to life or limb \citep{Peters2009} are considered a part of this term.} Farmers in ``light insurgent activity areas'' are assigned a random base value between $f_1=$ \$0 and \$100 per Ha, those in ``substantial insurgent activity areas'' a value between $f_1=$ \$100 and \$200, and farmers in ``heavy insurgent activity areas'' a value between $f_1=$ \$200 and \$300. The maximum value of $f_1=$ \$300 per Ha is chosen because this is the amount counter-insurgency operations of the U.S. Marines have paid farmers to destroy their opium crops \citep{Chisholm2010}. We assume regions with high insurgency levels will be able to counter U.S. offers with incentives of all kinds whose total value is a similar dollar amount. 

The switching costs of farmer activity, $i_c$, were estimated to be \$500, \$1,875, and \$4,500 dollars per Ha for changing to $c=$ grain-cereal-fodder, poppy, and fruit-nut-vegetable respectively. This is equivalent to half of the revenue from one harvest. Because a farmer will choose the crop that will provide the most income for the following season, the switching costs affect the impact of interventions seeking to change farmer crop choices. A distinction between single and multi-year switching costs, which might arise for fruit-nut-vegetable crops was not included, but would not have any impact on the conclusions.

\begin{figure*}[tb]
\refstepcounter{figref}\label{exits}
\centerline{
\includegraphics[width=125mm]{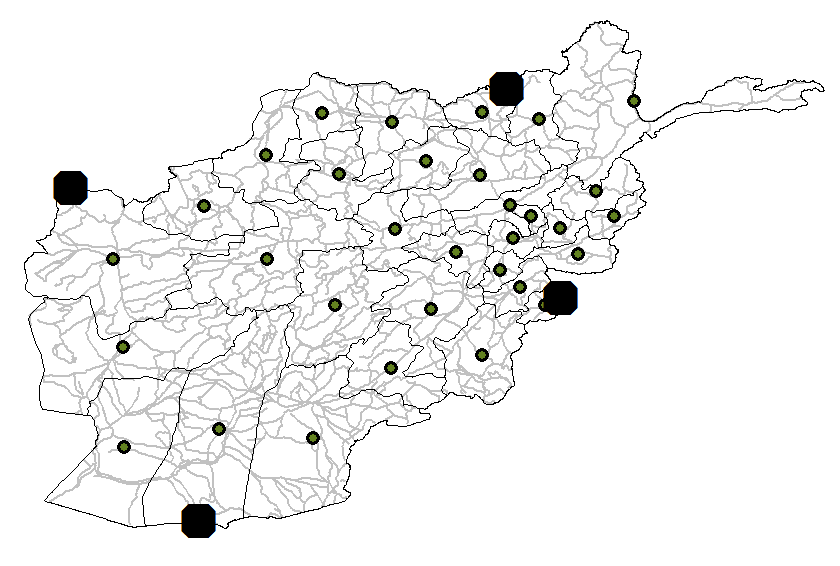}
}
\caption{Exit Points: Map showing the location of the four drug trafficking exit points (large black dots) considered in the model, Afghanistan province borders (dark grey lines) and their centroids (small dots) and the primary road networks (light grey lines).}
\caprule
\end{figure*}

Finally, the cost of trafficking opium, $e(\vect{r})$, is determined by calculating the road network distance from the centroid of a farmer's province to the nearest of four drug trafficking exit points (Figure \ref{exits}).{Only four exit points are represented in the model because it is infeasible to account for the myriad border crossings employed by traffickers. By using four points that are distributed along the border, the model captures the characterization of distances necessary for movement out of Afghanistan.} In the simulation, ``blockades'' corresponding to high levels of border security can be employed at any of the four exit points, requiring poppy farmers to send their crop to the closest exit point that is not blockaded, thus increasing trafficking costs. {Trafficking costs are designed to incorporate the effect of implementing blockades along certain regions of the Afghan border.}

The first step in calibrating the model is to develop a baseline scenario, where the number of poppy, grain-cereal-fodder, and fruit-nut-vegetable cells are consistent with the previously mentioned distribution of crops throughout Afghanistan. We find that an appropriate stable baseline scenario is obtained when the insurgency level, $f_0$, is held at the initial values, drug trafficking blockades are present at the northern two exit points, and the subsidy per hectare parameter, $s$, is set between \$1,100 and \$2,500. Within this range the subsidy parameter can be changed without affecting the system's behavior. The stability is due to the switching costs. We choose to use the minimum value in the subsidy range ($s=\$1,100$) as a reference value for the baseline scenario because we assume that no more than necessary would be spent to maintain stability. This is the principle of least effort, which follows from an assumption of short term rationality of policy makers who have competing uses for resources \citep{Zipf1949}. Increases in subsidy level that do not achieve any improvement are reversed, while decreases are maintained until they manifest deteriorating conditions. This results in the level of support being just above the minimum needed to sustain the current level of poppy growth. We find that the subsidy level assumptions are reasonable considering the present levels of international aid in Afghanistan. Providing a subsidy of \$1,100 per hectare to the farmers tending to Afghanistan's 5,700,000 hectares of arable land results in a total cost of around 6 billion dollars per year. This is a reasonable estimate of the security, infrastructure and more direct assistance to Afghani farmers, given that it is 5\% of the estimated US military expenditures of \$120 billion \citep{DeYoung2011} and on the order of the international humanitarian aid to Afghanistan of \$36 billion between 2001--2009 \citep{IRIN2010}, including \$4 billion spent by the US on aid in 2010 \citep{Cornwell2011}. 

It is important to note that the subsidy range that produces a stable scenario is dependent upon the assumptions made about switching costs and the blockade strategies. The present assumptions about border security are reasonable given the relatively high security in the northern parts of Afghanistan and the porous southern border with Pakistan \citep{Trofimov2010}. The stable subsidy range is sensitive to the switching cost values. Higher switching costs would expand and shift the stable subsidy range to higher values, while lower switching costs would contract and shift the range to lower values. The values chosen provide a useful illustration of the system behaviors. The following section describes a number of scenarios of policy interventions.

\section{Intervention scenario results}

The general behavior of the model can be understood from direct considerations. A sufficient increase in the subsidy of licit crops causes a reduction in the growth of poppy. Poppy production increases (decreases) with increased (decreased) levels of insurgency. Improved levels of border security, particularly with Pakistan, reduce the growth of poppy. A complete blockade of all border crossings eliminates poppy growth everywhere. Changes are inhibited by the cost associated with changing crops. Aside from extreme parameter values, land associated with fruit-nut-vegetable growth does not participate in the poppy economy, due to relatively high prices for those crops. These effects, however, interact with each other in specific ways that can be understood from simulations of the model. 
For example, subsidies must be raised beyond the baseline equilibrium subsidy range maximum of $s=$ \$2,500 per hectare in order to achieve a reduction in poppy cultivation. 
At reduced values of subsidy below \$1,100, poppy growth increases dramatically, by an amount that depends on the security scenario chosen. Thus, reducing the growth of poppy requires exceeding a certain threshold of economic investment and security enhancement. There is a range of values, which yield similar results. In that range, increasing investment in subsidy would increase costs but not achieve any significant impact. However, at the borders of that range, changes in subsidies move the system into a more or less desirable condition.

\begin{figure*}
\begin{leftfullpage}
\refstepcounter{figref}\label{3scen}
\centerline{
\includegraphics[width=170mm]{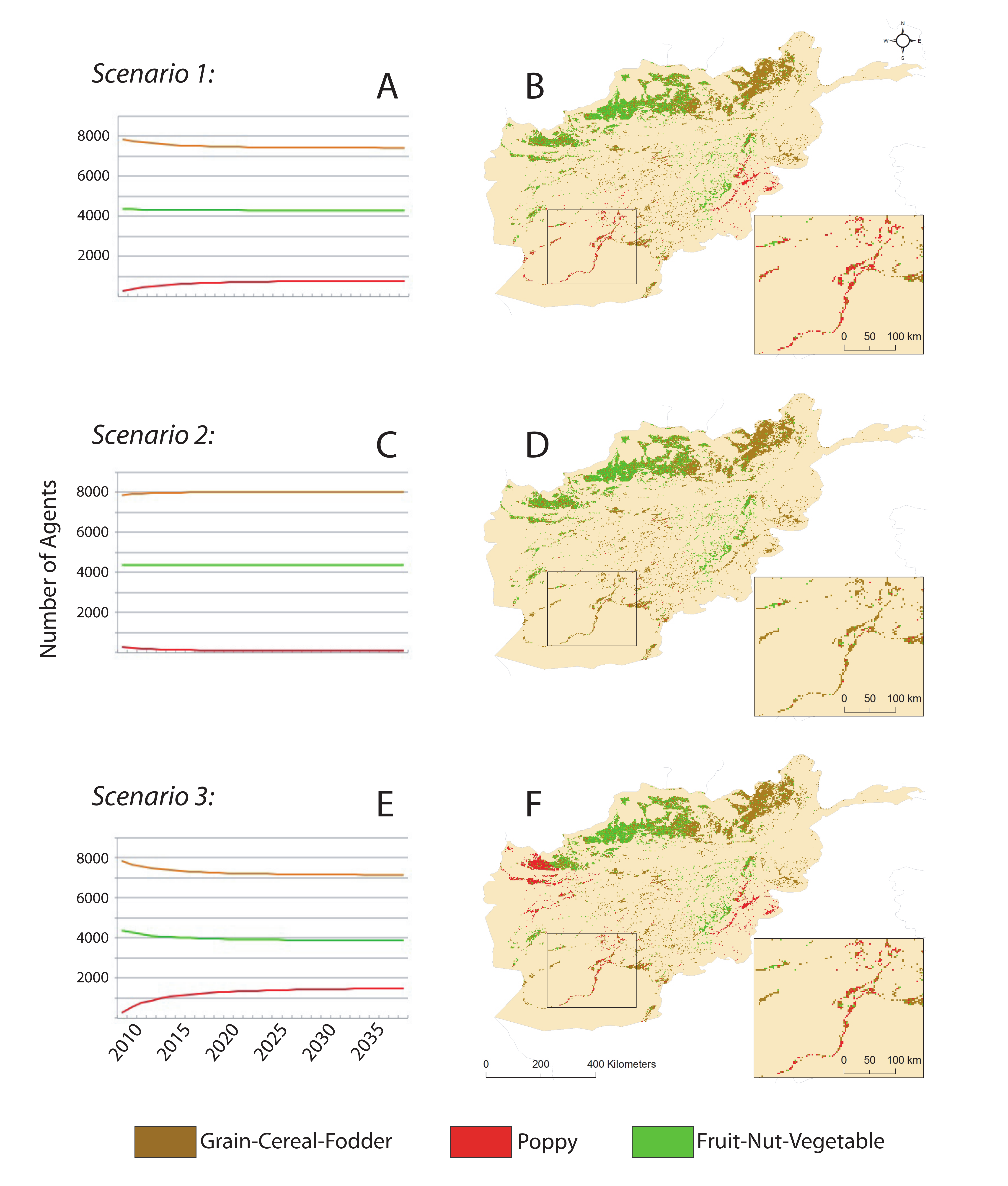}
}
\caption{\emph{(caption on following page)}\hfill}
\end{leftfullpage}
\end{figure*}

A few scenarios are shown in Figure \ref{3scen}. For each scenario, we present a plot of the number of agents farming the three types of crops at each each time step, where a time step represents a year. In addition, we present maps of the resulting stable geographic distributions of crops at the end of each simulation. The first scenario to be examined involves increasing the level of insurgent influence, $f_0$, holding all other parameters at their baseline values. A resurgence of insurgency throughout Afghanistan is likely to correspond to a concurrent increase in poppy crop production, as the funding of such a movement would benefit from illegitimate sources, such as drug trafficking, as seen in Figure \ref{3scen}A-B. The expansion of opium production occurs in the southwest and southeast of Afghanistan, where insurgency, $f_1$, is already relatively high, drug trafficking exit points are nearby, and most licit agriculture is of the less profitable grain-cereal-fodder variety.

\begin{figure*}[h]
\renewcommand\thefigure{\ref{3scen}}
\caption{\emph{(preceding page)} Time-series plots of number of agents farming each type of crop (A,C,E) and stable esimulation end-state (B,D,E) for each of the three experimental scenarios. Scenario 1 (A,B): Conditions of increased insurgent influence by an effective economic value of \$200. Other parameters are at baseline values (subsidy: \$1,100, drug trafficking blockades at the NE and NW exit points). Poppy cultivation spreads in the southeastern provinces of Afghanistan with the increase in insurgent influence and easy access to the SE exit point. Scenario 2: Conditions of increased licit crop subsidy to \$2,200 and a blockade placed at the southwestern drug trafficking exit point. In this scenario, most poppy growers in the southwestern provinces transition to grain-cereal-fodder crops without a resurgence of poppy growth elsewhere. Scenario 3: Conditions of an increase in insurgency influence with a financial value of \$200, a blockade placed at the southwestern drug trafficking exit point, and the removal of the blockade in the northwest. The removal of the northwestern blockade allows for easier trafficking out of the NW exit point. The increase in insurgent influence results in a number of farmers near the unblocked NW and SE exit points to transition to poppy.}
\caprule
\end{figure*}

The second scenario considers the impact of increasing the subsidy level from $s=$ \$1,100 to \$2,200 and implementing a drug trafficking blockade at the southwestern exit point. This scenario is designed to explore how a well-funded counter-narcotic campaign might eradicate opium production in the southwestern region of the country. Farmers who were initially growing poppy experience both incentives (subsidies) and disincentives (increased trafficking costs), resulting in the majority switching to their base crop. As seen in Figure \ref{3scen}C-D, such a strategy reduces the number of poppy crop cells to under 100 agents equivalent to a 62\% reduction of production.

The final scenario involves a slight increase of the insurgent influence parameter that adds \$200 to a farmer's baseline insurgent influence value, $f_1$ and the removal of the northwestern drug trafficking blockade. This scenario might be realized if international forces experience additional constraints on their financial and physical resources. We also considered the addition of a southwestern drug trafficking blockade in this scenario. Figure \ref{3scen}E-F shows that the increase in insurgent influence and removal of the drug trafficking blockade in the northwest result in increases in opium production in the northwest and southeast, the new blockade at the southwestern exit point does not have much, if any, effect on the poppy growth in the southwest. Social and economic forces in this region keep poppy production stable despite the local blockade.

\section{Conclusions}

The dynamic model presented in this paper captures abstracted, but important causal forces that influence Afghan farmers' decisions to grow poppy instead of licit crops. In the first and third scenarios discussed in the previous section, increases and decreases in poppy production in the northwest and southeast regions are caused by slight changes to the insurgent influence parameter. However, changes in poppy production in the southwestern region are more difficult to achieve. This is consistent with the present political and economic climate in southwestern provinces like Helmand, Kandahar, and Uruzgan. The model shows that a sustained effort of subsidizing farmers at a level near the upper bound of the baseline equilibrium range of \$1,100 to \$2,500 per hectare, while also targeting specific trafficking routes can rid southwestern Afghanistan of poppy without driving it to other regions. This reveals two important traits about the system at large. First, if subsidies are to be used as a policy lever in the fight against poppy growth, then an understanding of what subsidy values act as thresholds for desired shifts in the system is required. Secondly, blockade strategies that concentrate resources in targeted areas can have an effect across the entire region of interest.

The large required increase in investment to achieve improvements follows directly from the principle of minimum effort in the baseline ``status quo" scenario. Because policy decisions will tend to reduce support in the face of competing priorities, achieving a significant improvement requires an investment large enough to exceed the range of ineffective interventions. On the other hand any increase in insurgency or decrease in investment should immediately result in a deteriorating conditions. 

The importance of the trafficking blockades should be emphasized. The model demonstrates that boundary interventions (e.g. restricting trafficking opportunities), as opposed to internal interventions (e.g. increasing farmer subsidies), can be effective at achieving the goal of reducing the presence of poppy crops, but only if they are sufficient to blockade all major exit points (and combat the opening of others). Given that Afghanistan is the dominant supplier of the world's poppy, a well-implemented blockade will reduce the appeal and profits associated with the crop within the country. While power struggles across Afghanistan make a complete blockade of all parts of the border infeasible, policies that target border regions most suitable for trafficking poppy and opium out of Afghanistan can potentially disrupt the trade of this illicit crop. 

Implementing a thorough blockade, however, would have important social consequences for the large number of workers tied to the agricultural industry. While some poppy farmers may be able to switch to legal crops, additional opportunities should be provided to those who have no other means for making a living. While not included in our model, this can be reasonably expected to include those involved in the opium production and trafficking process. The implications of eliminating this portion of the Afghan economy could include an increase in social unrest due to unemployment and instability in the agricultural sector.

One limitation of the present implementation of this model is that crop prices are held constant. Because Afghanistan is responsible for the production of such a large portion of the world's opium supply, a reduction in the number of poppy farms would likely make the crop more profitable and create a balancing feedback, increasing the economic incentive for production and thus resisting the further reduction of poppy farming due to the higher prices. While a dynamic poppy pricing model might be an improvement, the effect of external price increases may be limited because much of the money gained from selling opium goes directly to warlords, drug traffickers and dealers, and government figures \citep{Goodson2005}. This shields poppy farmers from price swings to some extent, and makes the use of fixed prices in the simulation more reasonable as a first approximation \citep{UNODC2006}. Adding a more detailed description of current Afghan farms and farming, and the related socio-economic forces at work, may result in more precise analyses, however this should not affect the general trends described above. 

{Annual and progressive climate variations, which might also affect crop choices, would not affect the overall analysis of policy impacts or its conclusions.}

In this paper we have constructed a model that describes the role of economic forces in Afghanistan's farmers' choices. The Afghan farmer choice model uses geographic data to provide new policy insights into the dynamics behind the production of poppy crops. While our considerations represent only part of the worldwide opium trade network, understanding what actions are likely to reduce the number of farmers choosing to cultivate poppy may be helpful in constructing policy based upon achievable targets. This is particularly important since the model shows that without a certain threshold level of financial and/or security efforts, impact will be very limited. Through this understanding, more comprehensive and systemic strategies can be devised to fight opium production, trafficking, and use.

This work was supported in part by ONR under grant N000140910516 and AFOSR under grant FA9550-09-1-0324.

\end{document}